\documentclass[twocolumn,showpacs,preprintnumbers,amsmath,amssymb,prb]{revtex4}
\usepackage{graphicx,tabularx}
\usepackage{amssymb}
\usepackage{dcolumn}
\usepackage[mathcal]{euscript}
\newcommand{\m}{\textrm{MA\lowercase{l}H}_4}
\newcommand{\md}{\textrm{M(A\lowercase{l}H}_4\textrm{)}_2}

\newcommand{\na}{\textrm{N\lowercase{a}A\lowercase{l}H}_4}
\newcommand{\mg}{\textrm{M\lowercase{g}(A\lowercase{l}H}_4\textrm{)}_2}
\newcommand{\mgh}{\textrm{MgH}_2}
\newcommand{\Am}{\AA $\;$}
\begin{document}
\title{\em Ab initio \em study of magnesium alanate, $\mg$}
\author{M. J. van Setten}
\affiliation{Electronic Structure of Materials, Institute for Molecules and Materials, Faculty
of Science, Radboud University Nijmegen, Toernooiveld 1, 6525 ED Nijmegen, The Netherlands}
\author{V. A. Popa}
\affiliation{Computational Materials Science, Faculty of Science and Technology and MESA+
Research Institute, University of Twente, P.O. Box 217, 7500 AE Enschede, The Netherlands}
\author{G. A. de Wijs}
\affiliation{Electronic Structure of Materials, Institute for Molecules and Materials, Faculty
of Science, Radboud University Nijmegen, Toernooiveld 1, 6525 ED Nijmegen, The Netherlands}
\author{G. Brocks}
\affiliation{Computational Materials Science, Faculty of Science and Technology and MESA+
Research Institute, University of Twente, P.O. Box 217, 7500 AE Enschede, The Netherlands}

\date{\today}

%\keywords{$\mg$, hydrogen storage, DFT, GW, enthalpy of formation, electronic structure}

\begin{abstract}
Magnesium alanate $\mg$ has recently raised interest as a potential material for hydrogen
storage. We apply \textit{ab initio} calculations to characterize structural, electronic and
energetic properties of $\mg$. Density functional theory calculations within the generalized
gradient approximation (GGA) are used to optimize the geometry and obtain the electronic
structure. The latter is also studied by quasi-particle calculations at the $GW$ level. $\mg$
is a large band gap insulator with a fundamental band gap of 6.5 eV. The hydrogen atoms are
bonded in AlH$_4$ complexes, whose states dominate both the valence and the conduction bands.
On the basis of total energies, the formation enthalpy of $\mg$ with respect to bulk
magnesium, bulk aluminum and hydrogen gas is $0.17$~eV/H$_2$ (at $T=0$). Including
corrections due to the zero point vibrations of the hydrogen atoms this number decreases to
$0.10$~eV/H$_2$. The enthalpy of the dehydrogenation reaction
$\mg\rightarrow\mgh+2\textrm{Al}+3\textrm{H}_2(g)$ is close to zero, which impairs the
potential usefulness of magnesium alanate as a hydrogen storage material.
\end{abstract}

%----------------------------------------------------------------
\maketitle
%-----------------------------------------------------------------

The interest in alanate compounds as hydrogen storage materials was recently rekindled as the
kinetics of hydrogen adsorption/desorption improved dramatically by the addition of
transition metal catalysts.\cite{bog,bog_rev} Alanates $\m$, with M a lightweight alkaline
metal, have a high gravimetric hydrogen density, which is essential for their application as
storage materials. Most attention up to now has gone to sodium alanate, $\na$, which has a
hydrogen capacity of 7.5 wt.\ \%.\cite{bog_rev,bog2,chou} It releases hydrogen in a two stage
process. The two stages involve reaction enthalpies that are sufficiently small to be of
interest, namely 0.38 and 0.34~eV per H$_2$ molecule respectively. However, only three out of four
hydrogen atoms are released in this process, which lowers the effective hydrogen
storage capacity. This has stimulated the search for other suitable alanates.

Alanates $\md$, with M a lightweight alkaline earth metal, have an even higher gravimetric
hydrogen density. Recent interest turned to magnesium alanate, $\mg$, which has a hydrogen
capacity of 9.3 wt.\ \%.\cite{fich1,fich2,fich,foss} Upon heating $\mg$ releases hydrogen
according to the reaction
\begin{equation}
\label{desorption} \mg  \rightarrow \mgh + 2\textrm{Al} + 3\textrm{H}_2(g).
\end{equation}
Since decomposition of $\mgh$ takes place at too high a temperature to be of practical
use,\cite{bog_rev,bohm} it is the amount of hydrogen released in ({\ref{desorption}) that
determines the actual storage capacity of $\mg$. Still, this relatively large amount of 7.0
wt.\ \% makes magnesium alanate a good candidate for hydrogen storage. Only little is known
about the thermodynamics of this material, however.\cite{clau} Since up to now its
synthesis has proceeded via an indirect route, the first question is whether $\mg$ is
thermodynamically stable with respect to decomposition into its elements. The answer to this
question is relevant in the search for a more direct synthesis route.

A second question concerns the reaction enthalpy of ({\ref{desorption}). The ideal hydrogen
storage material should produce a hydrogen pressure of $0.1$~MPa at room temperature. The entropy
contribution of hydrogen gas at this temperature favors the right-hand side of
({\ref{desorption}). At $T=0$ the hydrogen desorption reaction should therefore have an
enthalpy of $\sim 0.4$~eV per desorbed H$_2$ molecule.\cite{bog_rev} Furthermore, the kinetics of
hydrogen adsorption/desorption should be sufficiently fast. Finding ways of improving the kinetics
can start from understanding the bonding in $\mg$, which is determined by the electronic
structure of the material.

In this paper we report the results of an \em ab initio \em study on the properties of
magnesium alanate. The structure is optimized and the electronic structure is calculated. We
characterize the bonding in $\mg$ and calculate the enthalpy of formation from its elements,
as well as the enthalpy of the dehydrogenation reaction ({\ref{desorption}).

\textit{Computational methods.} Total energies are calculated within density functional
theory (DFT), using the PW91 generalized gradient approximation (GGA) functional.\cite{gga}
We use the projector augmented wave (PAW) method\cite{paw,blo} and a plane wave basis set, as
implemented in the Vienna \em Ab initio \em Simulation Package
(VASP).\cite{vasp1,vasp2,vasp3} The atomic positions and the cell parameters, including the
cell volume, are optimized by minimizing the forces and stresses. A $7\times 7 \times 7$
Monkhorst-Pack $\bf k \rm$-point mesh is used for sampling the Brillouin zone.\cite{monk} A
kinetic energy cutoff of $312$~eV is used for the plane wave basis set. The reaction
enthalpies are calculated using a higher kinetic energy cutoff of $700$~eV to ensure
convergence.

If we calculate reaction enthalpies from total energy differences only, we neglect the
contributions from atomic vibrations. Such contributions are negligible for heavy
elements, whereas they may be significant for hydrogen. For each compound involved in the reaction we
calculate its zero point vibrational energy (ZPVE), $\frac{1}{2}\hbar\sum_j \omega_{j}$,
resulting from the vibrational modes $j$ in the optimized structure. Vibrational frequencies
$\omega_{j}$ are generated from a dynamical matrix, whose matrix elements (i.e., the force
constants) are calculated using a finite difference method.\cite{kress} For the hydrogen
molecules we calculate a ZPVE of 0.29~eV, in good agreement with the value of 0.27~eV
obtained from the experimental frequency.\cite{huber} We also consider the zero point
rotational energy (ZPRE) of the hydrogen molecules. Assuming that ortho- and para-hydrogen
are produced in a proportion of three to one, the average ZPRE of a hydrogen molecule is 0.011~eV, using the energy levels given in Ref.~\onlinecite{huber}. In summary, the reaction
enthalpies $\Delta H$ are calculated from
\begin{equation}
\label{enthalpy} \Delta H = \sum_{f}\left(E^\mathrm{tot}_f + E^\mathrm{ZPVE}_f\right) +
E^\mathrm{ZPRE}_{H_2}
 - \sum_{i}\left(E^\mathrm{tot}_i + E^\mathrm{ZPVE}_i\right)
\end{equation}
where $E^\mathrm{tot}_{f/i}$ denotes the total electronic energy of the reaction products $f$
or reactants $i$, $E^\mathrm{ZPVE}_{f/i}$ are the corresponding ZPVEs, and $E^{ZPRE}_{H_2}$
is the ZPRE of all hydrogen molecules involved in the reaction.

DFT calculations using the common density functionals give adequate values for ground state
properties such as total energies and vibrational frequencies. Excited state properties are
not given accurately, e.g., the electronic band gap is typically underestimated by $\sim 50
\%$. This in fact stems from an unjustified interpretation of the Kohn-Sham eigenvalues of
DFT as excitation energies. To calculate single-particle excitation energies, one should
solve a quasi-particle equation using the non-local, energy dependent self-energy. The $GW$
technique approximates the self-energy by a dynamically screened exchange interaction.
Constructing this interaction from the orbitals and eigenvalues obtained in a DFT calculation
with the local density approximation (LDA) is called the $G_0W_0$ approximation. It leads to
accurate band structures and band gaps for a wide range of semiconductors and
insulators.\cite{aryasetiawan} $GW$ calculations have also been successfully applied to metal
hydrides.\cite{gelderen1,gelderen2}

We start from an LDA calculation using norm conserving pseudo potentials and a plane wave
kinetic energy cutoff of 748~eV.\cite{TM} The screened interaction $G_0W_0$ is then
calculated using the real space, imaginary time formalism.\cite{gelderen2,rieger} For these
calculations we use 150 LDA states, a $13\times 13\times 19$ real space grid sampling of the
unit cell, and an interaction cell consisting of $5\times 5\times 4$ unit cells. The
quasi-particle equation is solved while neglecting the off-diagonal elements of the
self-energy between the LDA states. We estimate that the quasi-particle band gap of $\mg$ is
numerically converged to within $\pm 0.06$~eV.

\textit{Crystal structure.} Magnesium alanate has a CdI$_2$ structure with the Mg atoms on
the Cd positions and AlH$_4$ tetrahedra on the I positions; the space group is P$\bar{3}$m1
(164).\cite{fich} The structure basically consists of AlH$_4$ tetrahedra which form close packed
double layers perpendicular to the c-axis, alternated with a layer of Mg atoms, as shown in
Fig.~\ref{structure}.

Starting from the experimental structure proposed in Ref.~\onlinecite{fich}, we optimize the
atomic positions and cell parameters. As it turns out, for unit cell volumes in the range
125-150~\AA$^3$ the total energy only weakly depends upon the volume. We map out the total
energy as a function of the cell volume. At each volume we optimize the atomic positions and
the cell shape, and allow for breaking the symmetry. Interpolating this energy versus volume
curve gives a minimum energy at a cell volume of $143.26$~\AA$^3$. Optimizing the structure
at this volume gives the final results shown in Table~\ref{pos}. The calculated structure has
P$\bar{3}$m1 symmetry and is in good agreement with the recently obtained experimental
structure extracted from X-ray and neutron powder diffraction data.\cite{foss}

\begin{figure}[!tbp]
\centering
\includegraphics[width=5.0cm,keepaspectratio=true]{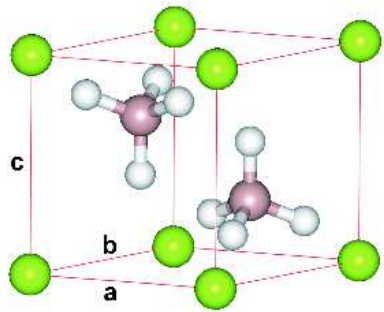}
\caption{(Color online) Crystal structure of $\mg$; space group
P$\bar{3}$m1}\label{structure}
\end{figure}

\begin{table}[!tbp]
\caption{Optimized crystal structure of $\mg$, compared to the experimental structure at 8~K
from Ref.~\onlinecite{foss}.\label{pos}}
\begin{ruledtabular}
\begin{tabular}{llccc}
Cell parameters && a & b & c\\
 &calc.& 5.23 & 5.23 & 6.04\\
 &exp.&5.21&5.21&5.84\\
\hline
Wyckoff positions& & x & y & z\\
Mg (1a) &calc.& 0 & 0 & 0    \\
&exp.&0&0&0\\
Al (2d) &calc.& 1/3 & 2/3 & 0.706\\
&exp.& 1/3 & 2/3 & 0.699\\
H1 (2d) &calc. & 1/3 & 2/3 & 0.442\\
&exp.& 1/3 & 2/3 & 0.424\\
H2 (6i) &calc. & 0.168 & $-$0.168 & 0.812 \\
&exp.& 0.167 & $-$0.167 & 0.811\\
\end{tabular}
\end{ruledtabular}
\end{table}

The Al-Al nearest neighbor distance within a double layer is 3.91~\AA , whereas the shortest
Al-Al distance between two double layers is 4.66~\AA . Mg atoms occupy octahedral
interstitial sites between two double layers with Mg-Al distances of 3.50~\AA. The AlH$_4$
tetrahedra are slightly distorted, but they retain a threefold rotation axis parallel to the
$\mathbf{c}$-axis. The Al-H1 and Al-H2 bond lengths are 1.60 and 1.62~\AA, and the H1-Al-H2
and H2-Al-H2 bond angles are $113.0^\mathrm{o}$ and $105.8^\mathrm{o}$. The geometry of the
AlH$_4$ tetrahedra is similar to that found in isolated (AlH$_4$)$^-$ ions, where the Al and
H atoms are covalently bonded.\cite{pullumbi} This geometry is quite different from those
found in neutral AlH$_x$ clusters.\cite{rao} The minimum Al-H and H-H distances between atoms
of different AlH$_4$ tetrahedra is 3.14~\Am and 2.63~\Am respectively, indicating that the
AlH$_4$ tetrahedra are clearly separated. The Mg atoms are octahedrally coordinated by H2
atoms with a Mg-H distance of 1.89~\Am and H-Mg-H angles of $86.9^\mathrm{o}$ and
$93.1^\mathrm{o}$. This coordination is not unlike that found in MgH$_2$.

\textit{Electronic structure.} Fig.~\ref{dos}(a) shows the electronic density of states (DOS)
of magnesium alanate obtained from the DFT/GGA calculation. It can be compared to the
calculated DOS of the lattice of (AlH$_4$)$^-$ tetrahedra, shown in Fig.~\ref{dos}(b). Here
the Mg ions have been removed and replaced by a homogeneous positive background charge. The
similarities between Figs.~\ref{dos}(a) and (b) demonstrate that the (AlH$_4$)$^-$ ions
strongly contribute to both the valence and the conduction bands of $\mg$. Such a dominance
of the anion is also observed in the simple ionic compound NaCl.\cite{boer1,boer2}

Projecting the valence states on atomic orbitals shows that Al and H contribute a comparable
amount, which is a strong indication for covalent bonding within the (AlH$_4$)$^-$
tetrahedra. The splitting into two valence bands, as is most clearly visible in
Fig.~\ref{dos}(b), is a remnant of the splitting between states of $s$-like ($A_1$) and
$p$-like ($T_2$) symmetry in a single tetrahedron. In an isolated (AlH$_4$)$^-$ ion, the
$sp$-gap is $\sim 4$ eV. This gap is closed to a certain extent by the interaction between
the (AlH$_4$)$^-$ ions, which results in a band dispersion of 2-3~eV. It shows that, although
the interaction between the (AlH$_4$)$^-$ tetrahedra is not negligible, it is weaker than the
interaction within a single tetrahedron.

If we compare the valence bands of Figs.~\ref{dos}(a) and (b) in more detail, we observe a
small peak in the DOS of $\mg$, which occurs within the $sp$-gap mentioned above. This peak
results from the hybridization between H and Mg $s$ states. Hybridization with Mg $p$ and $d$
states also gives less clearly visible contributions at higher energy. In any case, the H-Mg
hybridization is much weaker than the H-Al hybridization.

%-----dos
\begin{figure}[!tbp]
\centering
\includegraphics[angle=270,width=8.0cm,keepaspectratio=true]{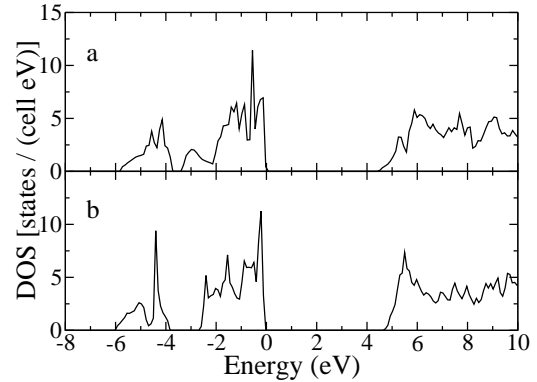}
\caption{(a) Electronic density of states (DOS) of $\mg$. The zero of energy is at the top of
the valence band. (b) DOS of (AlH$_4^-$)$_2$ with a positive homogenous background
charge.}\label{dos}
\end{figure}

The DFT/GGA band structure of $\mg$ is shown in Fig.~\ref{bands} It has an indirect band gap
of 4.4~eV. The bottom of the conduction band is at A and the top of the valence band located at maximum of AH at 0.7AH, $\left[\textrm{A}=\left( 0, 0, \frac12
\right),\;\textrm{H}=\left(\frac{1}{3},\frac{1}{3},\frac{1}{2}\right)\right]$. The band gap
as obtained from the $G_0W_0$ calculation is 6.5 eV. This classifies $\mg$ as a large band
gap insulator, which is typical for ionic compounds. The dispersions of the highest valence
and the lowest conduction bands in a direction along the $\mathbf{c}$-axis are rather small.
The direct band gap at A is 6.9~eV; the direct band gap at $\Gamma$ is 6.9~eV. More
generally the dispersion of the $G_0W_0$ bands is very similar to that of the GGA bands. The
total band width, 6.0~eV, of the $G_0W_0$ valence bands is $\sim 0.1$~eV larger than the
the GGA valence band width. 

The layered structure of $\mg$, see Fig.~\ref{structure}, does not imply that the
interactions in the compound are strongly anisotropic. The dispersions of the bands in
various directions are similar, compare e.g. the $\Gamma$A and the $\Gamma$M directions
$\left[\textrm{M}=\left( \frac12, 0, 0 \right)\right]$. This indicates that the interactions
between the ions within a layer (the $\mathbf{ab}$-plane) are comparable to those
perpendicular to the layers.

%-----band structure
\begin{figure}[!tbp]
\centering
\includegraphics[angle=270,width=8.0cm,keepaspectratio=true]{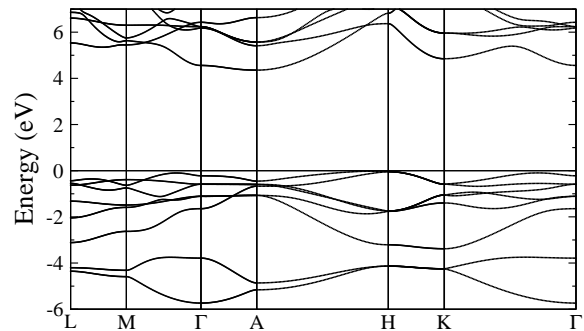}
\caption{DFT/GGA band structure of $\mg$. The zero of energy is at the top of the valence
band.} \label{bands}
\end{figure}

\textit{Formation enthalpy.} The formation enthalpy of $\mg$ with respect to its elements
corresponds to the enthalpy of the reaction
\begin{equation}
\label{formation} \mg \rightarrow \textrm{Mg} + 2\textrm{Al} + 4\textrm{H}_2(g).
\end{equation}
Here Mg and Al are in the crystalline phase, whereas H$_2$ is in the gaseous phase. For
aluminum we use the fcc structure with a lattice parameter of $4.05$~\Am and for magnesium we used
the hcp structure with lattice parameters $a=3.21$~\AA, $c=5.21$~\AA. The total energies and
ZPE corrections are given in Table~\ref{energies}. From these numbers, the reaction enthalpies
are then calculated using Eq.~(\ref{enthalpy}).

\begin{table}[!tbp]
\caption{Total energies (with respect to non spin polarized model atoms), zero point
vibrational energies (ZPVE) and zero point rotational energy (ZPRE) in eV/formula
unit.}\label{energies}
\begin{ruledtabular}
\begin{tabular}{lrrr}
       & E$^{\textrm{TOT}}$ & E$^{\textrm{ZPVE}}$ & E$^{\textrm{ZPRE}}$\\
\hline
Mg     &  $-1.524$ &  $0.001$ &\\
Al     &  $-3.698$ &  $0.004$ &\\
H$_2$  &  $-6.792$ &  $0.294$ & $0.011$\\
$\mgh$ &  $-8.983$ &  $0.402$ &\\
$\mg$  & $-36.764$ &  $1.520$ &\\
\end{tabular}
\end{ruledtabular}
\end{table}

The reaction enthalpy of (\ref{formation}) is $0.17$~eV per H$_2$ molecule on the basis of
total energies. If we include the ZPE, the reaction enthalpy decreases to $0.10$~eV/H$_2$.
Since the formation enthalpy is positive, it should, in principle, be possible to synthesize
$\mg$ from the elements. Note that at this energy scale, the contributions due to the zero
point motions of the hydrogen atoms are not negligible. In general, they tend to make a
negative contribution to the formation enthalpy of the metal hydride, since the motion of a
hydrogen atom in the crystal is more confined than in the gas phase.

To calculate the reaction enthalpy of (\ref{desorption}) one also needs the optimized
structure and ZPVE of $\mgh$. $\mgh$ has space group $P4_2/mnm$ (136) and its calculated
lattice parameters are $a=4.51$~\Am and $c=3.01$~\AA. The magnesium and hydrogen atoms are at
the $2a$ and $4f$ $(x=0.304)$ Wyckoff positions, respectively. As a check on the accuracy of
these calculations we can also extract the formation enthalpy of $\mgh$ with respect to its
elements
\begin{equation}
\label{formationmgh} \mgh \rightarrow \textrm{Mg} + \textrm{H}_2(g).
\end{equation}
The reaction enthalpy of (\ref{formationmgh}) is $0.67$~eV/H$_2$ without ZPE corrections and
$0.57$~eV/H$_2$ with ZPE corrections. This value is in reasonable agreement with the value of
$0.76$~eV/H$_2$, which is extracted by extrapolating the experimental results to zero
temperature.\cite{bohm,griess}

The calculated reaction enthalpy of (\ref{desorption}) is $0.003$~eV per H$_2$ molecule in
the gas phase, without ZPE correction. This is negligibly small, but consistent with earlier
experimental data.\cite{clau} Moreover, including the ZPE correction makes the reaction
enthalpy actually slightly negative, i.e. $-0.06$~eV/H$_2$. In any case this number is
significantly less than the $\sim 0.4$ eV/H$_2$ which, based on thermodynamics, is required to
make $\mg$ a good material for hydrogen storage. Further investigations are needed to see
whether e.g. alloying would increase the stability of magnesium alanate.

%----------------------------------------------------------------------

\textit{Acknowledgments.} We thank Prof. Dr. R. A. de Groot, Prof. Dr. P. J. Kelly, and Dr.
B. Dam for helpful discussions. This work is part of the research programs of `Advanced
Chemical Technologies for Sustainability (ACTS)' and `Stichting voor Fundamenteel Onderzoek
der Materie (FOM)', financially supported by `Nederlandse Organisatie voor Wetenschappelijk
Onderzoek (NWO)'.

%----------------------------------------------------------------------

\end{document}